\documentclass[journal]{IEEEtran}
\usepackage{enumerate}
\usepackage{cite}
\usepackage{graphicx}
\usepackage{alltt}
\usepackage{amsmath}
\usepackage{amsopn}
\usepackage{amssymb}
\usepackage[table]{xcolor}

\newtheorem{lemma}{Lemma}
\newtheorem{proposition}{Proposition}

\newtheorem{claim}{Claim}

\DeclareMathOperator{\deter}{det}

\DeclareMathOperator{\GF}{GF}

\title{Locally Repairable Regenerating Code Constructions}

\author{Imad~Ahmad,~\IEEEmembership{Student Member,~IEEE,}
and~Chih-Chun~Wang,~\IEEEmembership{Senior Member,~IEEE}
\thanks{This work was supported in parts by NSF grants CCF-0845968, CNS-0905331, CCF-1422997, and ECCS-1407604.}

\thanks{I. Ahmad and C.-C. Wang are with the School of Electrical and Computer Engineering, Purdue University, West Lafayette,
IN, 47906 USA e-mail: \{ahmadi,chihw\}@purdue.edu.}

}
\begin{document}
\maketitle
\thispagestyle{plain}
\pagestyle{plain}

\begin{abstract}
In this work, we give locally repairable regenerating code (LRRC) \cite{arxiv1,arxiv2,arxiv_multiple,ahmad2015locally,hollmann2014minimum} constructions that can protect the file size promised by the graph analysis of \emph{the modified family helper selection (MFHS) scheme} \cite{arxiv_multiple} at the minimum-bandwidth-regenerating (MBR) point. 
\end{abstract}

\section{Introduction}
The aim of this work is to provide code constructions of locally repairable regenerating codes (LRRCs) \cite{arxiv1,arxiv2,arxiv_multiple,hollmann2014minimum} that achieve the minimum-bandwidth-regenerating (MBR) point of the modified family helper selection (MFHS) scheme \cite{arxiv_multiple}. It is worth mentioning that related existing works to LRRCs can be found in \cite{rawat2014optimal,kamath2014codes} and also see the references in our previous work \cite{ahmad2014when, arxiv1,arxiv2,arxiv_multiple,ahmad2015locally}.

\section{Random Linear Code Construction for a Class of $(n,k,d,r)$ Parameters}\label{sec:rand_ex}
\par In this section, we prove the existence and construction of linear locally repairable regenerating codes (LRRCs) that achieve the MBR point of the MFHS scheme for $(n,k,d,r)$ values that satisfy that $n-d-r=2$ and $n\bmod (n-d-r)=0$, i.e., the family size in the MFHS scheme is 2 and there are no incomplete families. The code existence proof idea and construction are inspired by the work in \cite{wu2010existence}. 

\par The goal of this section is to prove the existence and construction of linear codes that can protect a file of size
\begin{align}\label{eq:file_size}
\mathcal{M}=\sum_{i=1}^k (d-y_i(\pi_f^*)) \text{ packets}
\end{align}
against any $(n-k)$ simultaneous failures.

\par We start first by describing the notation that will be used in this section. We represent the original file by an $\mathcal{M}\times W$ matrix $\mathbf{X}$ defined over finite field $\GF(q)$ denoted by $\mathbb{F}$, where $q$ is a fixed finite field size satisfying $q>nd\mathcal{M}|\mathcal{H}|$ and $\mathcal{H}$ is a finite set that will be defined shortly. The file is viewed in the following as containing $\mathcal{M}$ packets, where the packet is thought of as the smallest unit of data. Recall that we are considering the MBR point in this section. Therefore, the file size is $\mathcal{M}$ packets, the storage-per-node $\alpha=d$ packets and the repair-bandwidth per-helper $\beta=1$ packet. Now, storage node $i$ stores $\alpha=d$ packets $\mathbf{X}^{T}\mathbf{Q}_i$, where $\mathbf{Q}_i$ is an $\mathcal{M}\times d$ matrix. By this definition, matrices $\mathbf{Q}_1,\mathbf{Q}_2,\dots,\mathbf{Q}_n$ are sufficient to completely specify a code.

\par As in \cite{wu2010existence} and \cite{koetter2003algebraic}, the existence proof is based on applying the Schwartz-Zippel theorem \cite{motwani2010randomized}.

\begin{lemma}
Let $Q(x_1,\dots,x_n)\in \mathbb{F}[x_1,\dots,x_2]$ be a multivariate polynomial of total degree $d_0$ (the total degree is the maximum degree of the additive terms and the degree of a term is the sum of exponents of the variables). Fix any finite set $\mathbb{S}\subseteq \mathbb{F}$, and  let $r_1,\dots,r_n$ be chosen independently and uniformly at random from $\mathbb{S}$. Then if $Q(x_1,\dots,x_n)$ is not equal to a zero polynomail,
\begin{align}
\text{Pr}[Q(r_1,\dots,r_n)=0]\leq\frac{d_0}{|\mathbb{S}|}.
\end{align}
\end{lemma}

As in \cite{wu2010existence} and \cite{koetter2003algebraic}, the core idea of the proof relies on using induction to represent the dynamics of the storage problem where the induction invariant is the condition that guarantees the existence of the code. Then, the induction invariant is formulated as a product of multivariate polynomials where each polynomial is shown to be non-zero to complete the proof by invoking the Schwartz-Zippel theorem. The existence proof itself implies that we can construct linear codes \emph{randomly}.

\par Before we give the induction invariant and show how the induction process works, we define an important set of vectors that we call set $\mathcal{H}$. We abuse the above notation and we let $\pi$ be a permutation of the storage node vector $(1,2,\dots,n)$ and $\pi(i)$ be its $i$-th coordinate value. Define vector $\mathbf{b}(\pi)$ such that its $i$-th coordinate $b_i(\pi)=(d-z_i(\pi))^+$, where $z_i(\cdot)$ is as defined in the proof of \cite[Proposition~12]{arxiv_multiple}. Moreover, we define vector $\mathbf{c}(\pi)$ as the truncated version of $\mathbf{b}(\pi)$, where by truncation we mean the following: (ii) we find the smallest $m$ such that $\sum_{i=1}^m b_i(\pi)\geq \mathcal{M}$; (ii) we set $c_i(\pi)=b_i(\pi)$ for $i=1$ to $(m-1)$. We then set $c_m(\pi)=\mathcal{M}-\sum_{i=1}^{m-1} b_i(\pi)$ and $c_i(\pi)=0$ for $i=(m+1)$ to $n$. To illustrate the constructions of $\mathbf{b}(\pi)$ and $\mathbf{c}(\pi)$, consider $(n,k,d,r)=(6,4,3,1)$ and $\pi=(2,3,4,1,5,6)$. We have by \eqref{eq:file_size} that $\mathcal{M}=7$ and we can get that $\mathbf{b}(\pi)=(3,2,2,1,0,0)$ by the definition of vector $\mathbf{b}(\pi)$. Truncating $\mathbf{b}(\pi)$ as described above, we get $\mathbf{c}(\pi)=(3,2,2,0,0,0)$. Now, we will give an important property of $\mathbf{b}(\pi)$ and $\mathbf{c}(\pi)$ in the following lemma.

\begin{lemma} \label{lem:b_c}
For $n-d-r=2$ and $n\bmod(n-d-r)=0$, we have that for any node permutation $\pi$, $\mathbf{b}(\pi)$ and $\mathbf{c}(\pi)$ satisfy that $b_i(\pi)\geq b_{i+1}(\pi)$ and $c_i(\pi)\geq c_{i+1}(\pi)$.
\end{lemma}

\begin{IEEEproof}
The proof is divided into two cases.
\par \emph{Case~1}: $FI(\pi(i))= FI(\pi(i+1))$. Since $(d-z_i(\pi))=(d-z_{i+1}(\pi))$, we have that $b_i(\pi)=b_{i+1}(\pi)$.
\par \emph{Case~2}: $FI(\pi(i))\neq FI(\pi(i+1))$. The value $(d-z_i(\pi))$ is smallest when all nodes $\pi(1)$ to $\pi(i-1)$ are not in the family $FI(\pi(i))$. On the other hand, the largest $(d-z_{i+1}(\pi))$ value is when the other node in family $FI(\pi(i+1))$ is in the nodes $\pi(1)$ to $\pi(i-1)$. Notice, however, that we have that $FI(\pi(i))\neq FI(\pi(i+1))$. This fact will decrease $(d-z_{i+1}(\pi))$ by a value of 1. Therefore, the smallest $(d-z_{i}(\pi))$ still satisfies $(d-z_i(\pi))\geq (d-z_{i+1}(\pi))$. Recall that $b_i(\pi)=(d-z_i(\pi))^+$ and $b_{i+1}(\pi)=(d-z_{i+1}(\pi))^+$ by definition. Therefore, $b_i(\pi)\geq b_{i+1}(\pi)$.

\par By the above two cases, we have proved that $b_i(\pi)\geq b_{i+1}(\pi)$ for any node permutation $\pi$. Since $\mathbf{c}(\pi)$ is a truncated version of $\mathbf{b}(\pi)$, $\mathbf{c}(\pi)$ also satisfies that $c_i(\pi)\geq c_{i+1}(\pi)$ for any permutation $\pi$. Hence, the proof of this lemma is complete.  
\end{IEEEproof}

\par We are now ready to define set $\mathcal{H}$ that will be used to write the induction invariant. Suppose we have two $n$-dimensional vectors $\mathbf{a}$ and $\mathbf{b}$. We say that $\mathbf{a}$ majorizes $\mathbf{b}$, denoted by $\mathbf{a}\succeq \mathbf{b}$, if $\sum_{i=1}^m a_{[i]} \geq \sum_{i=1}^m b_{[i]}$ for all $m=1$ to $n$, where $a_{[i]}$ and $b_{[i]}$ are the $i$-th largest values in $\mathbf{a}$ and $\mathbf{b}$, respectively.

\par Now, we define set $\mathcal{H}$ as follows. A vector $\mathbf{h}\in \mathcal{H}$ if it satisfies the following two conditions: (i) it is an $n$-dimensional integer vector such that $0\leq h_i\leq d$ for all $i=1$ to $n$ and (ii) there exists a node permutation $\pi$ such that $h_{\pi(i)}\geq h_{\pi(i+1)}$ for $i=1$ to $(n-1)$ and $ \mathbf{c}(\pi)\succeq\mathbf{h}$. Using set $\mathcal{H}$, we write the induction invariant as
\begin{align}\label{eq:induc_invar}
\prod_{\mathbf{h}\in \mathcal{H}} \deter ([\mathbf{Q}_1\mathbf{E}_{h_1}, \dots, \mathbf{Q}_n\mathbf{E}_{h_n}])\neq 0,
\end{align}
where $E_x$ denotes a selection matrix such that $\mathbf{Q}_i\mathbf{E}_x$ holds the first $x$ columns of $\mathbf{Q}_i$, and recall that $h_i$ is the $i$-th coordinate of vector $\mathbf{h}$.

\par In order to prove that there exists a code that can protect a file of size $\mathcal{M}$, it is sufficient to prove that initially we can get matrices $\mathbf{Q}_i$ such that the induction invariant \eqref{eq:induc_invar} is satisfied over $\mathbb{F}$ and then show that the induction invariant can be maintained after any node arbitrary failure/repair that can happen. This is sufficient because the reconstruction (MDS) property or condition is weaker than the condition in \eqref{eq:induc_invar}. Specifically, since all possible $\mathbf{c}(\pi)$ vectors belong to set $\mathcal{H}$, we have that every subset of $k$ matrices from $\{\mathbf{Q}_1,\dots,\mathbf{Q}_n\}$ is full rank or equivalently can recontruct the file.

\par The first step thus is to prove that \eqref{eq:induc_invar} is satisfied initially by $\mathbf{Q}_1,\dots,\mathbf{Q}_n$ over $\mathbb{F}$. Notice, that the left-hand side (LHS) of \eqref{eq:induc_invar} can be thought of as a multivariate polynomial, where the variables are the entries of $\mathbf{Q}_1,\dots,\mathbf{Q}_n$. Since the size of set $\mathcal{H}$ is finite, the degree of the polynomial in $\eqref{eq:induc_invar}$ is at most $d_1=nd\mathcal{M}|\mathcal{H}|$ and by Schwartz-Zippel theorem we can find entries of the matrices $\mathbf{Q}_1,\dots,\mathbf{Q}_n$ such that \eqref{eq:induc_invar} is since $|\mathbb{F}|=q>d_1$.

\par Now, we suppose that the induction invariant in \eqref{eq:induc_invar} is satisfied until stage $(t-1)$ over $\mathbb{F}$. Note that in order to be able to invoke Schwartz-Zippel theorem, the LHS of \eqref{eq:induc_invar} has to be a non-zero polynomial. It is not hard to see that for each vector $\mathbf{h}\in \mathcal{H}$, the determinant term in \eqref{eq:induc_invar} is a non-zero polynomial by using the column vectors of the $\mathcal{M}\times\mathcal{M}$ identity matrix and placing them on the first columns of $\mathbf{Q}_1,\dots,\mathbf{Q}_n$. Specifically, $\mathbf{Q}_1$ takes the first $h_1$ columns of the identity matrix, $\mathbf{Q}_2$ takes the first $h_2$ columns of the remaining columns and so on and so forth. Since the product of non-zero polynomials is a non-zero polynomial, the LHS of \eqref{eq:induc_invar} is a non-zero polynomial.

\par Without loss of generality, suppose node 1 fails and a newcomer replacing node 1 communicates with helpers $\{x_1,\dots,x_d\}\subset \overline{D}(1)$ for repair. By the nature of the repair problem and the fact that we are considering linear codes, we can write the coding matrix on the newcomer as 
\begin{align}
\mathbf{Q'}_1=[\mathbf{Q}_{x_1}\mathbf{b}_1,\dots,\mathbf{Q}_{x_d}\mathbf{b}_d]\mathbf{Z},
\end{align}
where $\mathbf{b}_i$ is a $d\times 1$ column vector and $\mathbf{Z}$ is a $d\times d$ matrix representing the possible linear transformation a newcomer can apply to the $d$ received repair packets.

\par In this step, we want to prove that it is possible to find $\mathbf{b}_1,\dots,\mathbf{b}_d$ vectors and a matrix $\mathbf{Z}$ over $\mathbb{F}$ that satisfy \eqref{eq:induc_invar}.

\par In total we have $2d^2$ variables. Therefore, the degree of the polynomial on the LHS of the new condition
\begin{align}\label{eq:induc_invar_new}
\prod_{\mathbf{h}\in \mathcal{H}} \deter ([\mathbf{Q'}_1\mathbf{E}_{h_1},\mathbf{Q}_2\mathbf{E}_{h_2}, \dots, \mathbf{Q}_n\mathbf{E}_{h_n}])\neq 0,
\end{align}
is at most $2d^2|\mathcal{H}|$. Notice that $2d^2|\mathcal{H}|\leq nd\mathcal{M}|\mathcal{H}|$ since $n\geq 2$ and $\mathcal{M}\geq d$. Therefore, since $q>nd\mathcal{M}|\mathcal{H}|$, by Schwartz-Zippel theorem, we have that we can find $\mathbf{b}_1,\dots,\mathbf{b}_n$ vectors and matrix $\mathbf{Z}$ over $\mathbb{F}$ such that \eqref{eq:induc_invar_new} is true. Assuming that the LHS of \eqref{eq:induc_invar_new} is non-zero. It turns out that proving this fact is non-trivial. The remainder of this section is dedicated to showing that the LHS of \eqref{eq:induc_invar_new} is a non-zero polynomial.

\par The proof idea for proving that the LHS of \eqref{eq:induc_invar_new} is a non-zero polynomial is as follows. We still suppose that node 1 fails without loss of generality and we suppose that node 1 will repair from helpers $\{x_1,\dots,x_d\}\subset \overline{D}(1)$. Recall that it is sufficient to prove that the polynomial of the determinant for each $\mathbf{h}\in \mathcal{H}$ is non-zero since the product of non-zero polynomials is a non-zero polynomial. Using this fact, we will consider any $\mathbf{h}\in \mathcal{H}$ first. Then, we will prove that we can always find a vector $\mathbf{h}'\in \mathcal{H}$ such that $h_1'=0$, and there exists a subset of nodes in $\{x_1,\dots,x_d\}$ of size $h_1$, $\{s_1,\dots,s_{h_1}\}$, such that $h_{s_1}',\dots,h_{s_{h_1}}'$ satisfy that $h_{s_i}'=h_{s_i}+1$ for $i=1$ to $h_1$. 

\par Since $\mathbf{h}'\in \mathcal{H}$, we have that
\begin{align}\label{eq:det_h_old}
\deter([\mathbf{Q}_2\mathbf{E}_{h_2'},\dots,\mathbf{Q}_n\mathbf{E}_{h_n'}])\neq 0.
\end{align}
 Now, we choose the vectors $\mathbf{b}_1,\dots,\mathbf{b}_d$ such that they select the $h'_{s_i}$-th column vector from each of $\mathbf{Q}_{s_1},\dots,\mathbf{Q}_{s_{h_1}}$ and we choose $\mathbf{Z}$ to be the $d\times d$ identity matrix. By \eqref{eq:det_h_old}, we have that 
\begin{align}
\deter([\mathbf{Q}_1'\mathbf{E}_{h_1} \mathbf{Q}_2\mathbf{E}_{h_2},\dots,\mathbf{Q}_n\mathbf{E}_{h_n}])\neq 0.
\end{align}
Since we have found one combination of the entries in the vectors $\mathbf{b}_1,\dots,\mathbf{b}_d$ and the matrix $\mathbf{Z}$ such that for the considered $\mathbf{h}\in \mathcal{H}$ the determinant is non-zero, then this polynomial is non-zero. Notice that we stated that we can find such $\mathbf{h}'\in \mathcal{H}$ for any $\mathbf{h}\in \mathcal{H}$. Therefore, the LHS of \eqref{eq:induc_invar_new} is a non-zero polynomial. Now, we are left with proving the fact that we can always find such $\mathbf{h}'\in\mathcal{H}$. In order to do that, we will need the following claim.

\begin{claim}\label{clm:swapping}
Consider any vector $\mathbf{h}\in \mathcal{H}$ where $n-d-r=2$ and $n\bmod(n-d-r)=0$. Consider a node permutation $\pi$ such that $h_{\pi(i)}\geq h_{\pi(i+1)}$. If $\mathbf{c}(\pi)\succeq \mathbf{h}$ and $h_{\pi(i)}=h_{\pi(i+1)}$ for some integer $i$, then $\mathbf{c}(\pi')\succeq \mathbf{h}$ where $\pi'$ is a node permutation such that $\pi'(i)=\pi(i+1)$, $\pi'(i+1)=\pi(i)$, and $\pi'(j)=\pi(j)$ for $j\in \{1,\dots,i-1,i+2,\dots,n\}$.
\end{claim}
\par The proof of this claim is relegated to Appendix~\ref{app:swapping_claim}.

\par We now describe procedure {\sc CONNECT} that takes the vector $\mathbf{h}\in \mathcal{H}$ as input and outputs a vector $\mathbf{h}'\in \mathcal{H}$ that satisfies the properties discussed above. After we describe procedure {\sc CONNECT}, we will prove the correctness of the procedure.

\par Now, we will define node classes that divide the nodes according to their values in $\mathbf{h}$. We have $(d+1)$ classes defined as $A_g=\{m:1\leq m\leq n, h_m=g\}$ for $g=0$ to $d$. Procedure {\sc CONNECT} is as follows:
\begin{enumerate}
\item Pick node permutation $\pi^{(0)}$ such that (i) $h_{\pi^{(0)}(i)}\geq h_{\pi^{(0)}(i+1)}$ and (ii) if $n_1\in A_g\cap \{x_1,\dots,x_d\}$ for some $g$ and $n_2\in A_g\backslash \{x_1,\dots,x_d\}$, then $p^{(0)}(n_1)<p^{(0)}(n_2)$, where $p^{(0)}$ satisfies $p^{(0)}(\pi^{(0)}(i))=i$.

\item Let $\mathbf{h}^{(0)}=\mathbf{h}$ and let $D^{(0)}=\{x_1,\dots,x_d\}$. Let $A_g^{(0)}=A_g$ for $g=0$ to $d$.
\par Note that jointly Claim~\ref{clm:swapping} and the fact that $\mathbf{h}\in \mathcal{H}$ imply that $\mathbf{c}(\pi^{(0)}) \succeq \mathbf{h}^{(0)}$.

\item Sequentially, do the following for $t=1$ to $h_1$:
\begin{enumerate}
\item Find $x\in D^{(t-1)}$ such that (i) $h_x^{(t-1)}\leq h_y^{(t-1)}$ for all $y\in D^{(t-1)}$ and (ii) $p^{(t-1)}(x)<p^{(t-1)}(y)$ for all $y\in A_{h_x}^{(t-1)}\cap D^{(t-1)}\backslash \{x\}$.
\item Let $h_x^{(t)}=h_x^{(t-1)}+1$, $h_1^{(t)}=h_1^{(t-1)}-1$, and $h_i^{(t)}=h_i^{(t-1)}$ for all $i\in \{2,\dots,x-1,x+1,\dots,n\}$.
\item Let $D^{(t)}=D^{(t-1)}\backslash \{x\}$ and $A_g^{(t)}=\{m:1\leq m\leq n, h_m^{(t)}=g\}$ for $g=0$ to $d$. Let $\pi^{(t)}$ such that (i) $p^{(t)}(\pi^{(t-1)}(i))<p^{(t)}(\pi^{(t-1)}(j))$ whenever $i<j$ and both $\pi^{(t-1)}(i)$ and $\pi^{(t-1)}(j)$ are not equal to 1, i.e., the order of the nodes in $\pi^{(t-1)}$ is preserved for all nodes except node 1, which can move to later position in the order (have larger $p$ value), and (ii) $\pi^{(t)}$ satisfies that $h^{(t)}_{\pi^{(t)}(i)}\geq h^{(t)}_{\pi^{(t)}(i+1)}$. This is always possible by the construction of $\pi^{(0)}$.
\end{enumerate}
\item Return the output $\mathbf{h}'=\mathbf{h}^{(h_1)}$.
\end{enumerate} 

\begin{proposition} \label{prop:connect_correct}
Procedure {\sc CONNECT} is correct.
\end{proposition}
\begin{IEEEproof} First, we need to show that the output of {\sc CONNECT} $\mathbf{h}'$ satisfies that $h_1'=0$ and $|\{h_i':h_i'=h_i+1,i\in\{x_1,\dots,x_d\}\}|=h_1$. We can see that by Step~3 of {\sc CONNECT}, $h_1$ always decreases by 1 with every iteration and with every iteration we are always choosing a helper in $\{x_1,\dots,x_d\}$ and adding 1 to its $h$ value. Since Step~3 is done exactly $h_1$ times, $\mathbf{h}'$ satisfies the properties. The other important property of $\mathbf{h}'$ that we need to show is that $\mathbf{h}'\in \mathcal{H}$. In order to do that, we need to prove two points (i) $0\leq h_i'\leq d$ (ii) there exists a node permutation $\pi$ such that $h_{\pi(i)}'\geq h_{\pi(i+1)}'$ and $\mathbf{c}(\pi)\succeq \mathbf{h}'$. 

\par Since $0\leq h_i \leq d$, then $h_i'>d$ is only possible for $i\in \{x_1,\dots,x_d\}$. Since in {\sc CONNECT}, only $h_1$ (and not all $d$ nodes) of the helpers $x_1,\dots,x_d$ will have their $h$ value incremented by 1, we need to have that at least $h_1$ of the helpers satisfy that their $h$ value is strictly less than $d$.

\par We will first argue that having more than two nodes of $\{x_1,\dots,x_d\}$ with an $h$ value equal to $d$ is not possible. The reason is as follows. Suppose we have more than two nodes like that. Let these nodes be $x_{s_1},\dots,x_{s_m}$. Then, construct a node permutation $\pi$ such that the first $m$ coordinates are $x_{s_1},\dots,x_{s_m}$ and $h_{\pi(i)}\geq h_{\pi(i+1)}$ for $i=2$ to $(n-1)$. Since $n-d-r=2$, the family size is 2 and $c_3(\pi)<d$. Thus, $\mathbf{c}(\pi)\nsucceq \mathbf{h}$ and this implies that $\mathbf{h}\notin \mathcal{H}$ by Claim~\ref{clm:swapping}. Therefore, by contradiction we have proved that we cannot have more than two nodes of $\{x_1,\dots,x_d\}$ with $h$ values equal to $d$. Now, if $h_1=d$, it is not possible to have any node of $\{x_1,\dots,x_d\}$ with $h$ value $d$ since that again means that $\mathbf{c}(\pi)\nsucceq \mathbf{h}$, where $\pi$ is the same as the $\pi$ we constructed above.

\par If $h_1=d-1$, it is not possible to have more than 1 node of $\{x_1,\dots,x_d\}$ with $h$ value $d$ for the same reason. In both cases, $0\leq h_i'\leq d$ for $i=1$ to $n$ since there are $h_1$ helpers with $h$ values strictly less than $d$. For $h_1\leq d-2$, since it is not possible to have more than 2 nodes of $\{x_1,\dots,x_d\}$ with $h$ values equal to $d$, then there are always $h_1$ helpers with $h$ values stricly less than $d$. Therefore, in all cases $\mathbf{h}'$ satisfies $0\leq h_i'\leq d$ for all $i=1$ to $n$.

\par At this point, we are left with proving that there exists a permutation $\pi$ such that $\mathbf{c}(\pi)\succeq \mathbf{h}'$. We will show shortly that $\pi^{(h_1)}$ is indeed this permutation. Stronger than that, we will prove that at the end of each iteration $t$ of {\sc CONNECT}, $\mathbf{c}(\pi^{(t)})\succeq \mathbf{h}^{(t)}$. The proof is by contradiction. Suppose that $\mathbf{c}(\pi^{(t)})\nsucceq \mathbf{h}^{(t)}$. Let $l$ be the smallest $l$ such that 
\begin{align} \label{eq:connect_notmaj}
h_{\pi^{(t)}(1)}^{(t)}+\dots+h_{\pi^{(t)}(l)}^{(t)}>c_1(\pi^{(t)})+\dots+c_l(\pi^{(t)}).
\end{align}
Notice that $l<k$. Let $l'$ be the largest integer such that $h_{\pi^{(t-1)}(l')}^{(t-1)}=h_{\pi^{(t-1)}(l)}^{(t-1)}$ and $\pi^{(t-1)}(l')\in D^{(t-1)}$, i.e., $l'$ is the largest integer such that $\pi^{(t-1)}(l')$ is a helper of $D^{(t-1)}$ in the node class $A_g^{(t-1)}$, where $g=h_{\pi^{(t-1)}(l)}^{(t-1)}$. We have the following claim based on the definition of $l'$.

\begin{claim}\label{clm:l_prime}
$h_{\pi^{(t)}(1)}^{(t)}+\dots+h_{\pi^{(t)}(l')}^{(t)}>c_1(\pi^{(t)})+\dots+c_{l'}(\pi^{(t)})$.
\end{claim}
We note that $l'\leq k$.
\begin{IEEEproof}
By the definition of $l$, we have that
\begin{align}
h_{\pi^{(t)}(1)}^{(t)}+\dots+h_{\pi^{(t)}(l-1)}^{(t)}\leq c_1(\pi^{(t)})+\dots+c_{l-1}(\pi^{(t)}).
\end{align}
We then get by \eqref{eq:connect_notmaj} that $h^{(t)}_{\pi^{(t)}(l)}>c_l(\pi^{(t)})$. Now, we have
\begin{align}
h_{\pi^{(t)}(1)}^{(t)}&+\dots+h_{\pi^{(t)}(l)}^{(t)}+h_{\pi^{(t)}(l+1)}^{(t)}+\dots+h_{\pi^{(t)}(l')}^{(t)}\nonumber\\
&> c_1(\pi^{(t)})+\dots+c_{l}(\pi^{(t)})+(l'-l)h_{\pi^{(t)}(l')}^{(t)}\label{eq:clm_l_prime_1}\\
&>c_1(\pi^{(t)})+\dots+c_{l}(\pi^{(t)})+(l'-l)c_l(\pi^{(t)})\label{eq:clm_l_prime_2}\\
&\geq c_1(\pi^{(t)})+\dots+c_{l}(\pi^{(t)})+c_{l+1}(\pi^{(t)})+\nonumber\\
&\quad \quad \dots+c_{l'}(\pi^{(t)})\label{eq:clm_l_prime_3},
\end{align}
where \eqref{eq:clm_l_prime_1} is by the fact that $h_{\pi^{(t)}(l)}^{(t)}=h_{\pi^{(t)}(l+1)}^{(t)}=\dots=h_{\pi^{(t)}(l')}^{(t)}$, \eqref{eq:clm_l_prime_2} by the fact that $h^{(t)}_{\pi^{(t)}(l)}>c_l(\pi^{(t)})$, and \eqref{eq:clm_l_prime_3} is by the fact that $c_i(\pi^{(t)})\geq c_{i+1}(\pi^{(t)})$. By \eqref{eq:clm_l_prime_3}, the proof of this claim is complete.
\end{IEEEproof}

\par Recall that we are proving that $\mathbf{h'}^{(t)}\in \mathcal{H}$ by contradiction. We have two cases:

\par \emph{Case~1}: if $1\in \{\pi^{(t)}(1),\dots,\pi^{(t)}(l')\}$. We have that
\begin{align}
h_{\pi^{(0)}(1)}^{(0)}+\dots+&h_{\pi^{(0)}(l')}^{(0)}\nonumber\\
&\geq h_{\pi^{(t)}(1)}^{(0)}+\dots+h_{\pi^{(t)}(l')}^{(0)}\nonumber\\
&\geq h_{\pi^{(t)}(1)}^{(t)}+\dots+h_{\pi^{(t)}(l')}^{(t)}\label{eq:connect_1}\\
&>c_1(\pi^{(t)})+\dots+c_{l'}(\pi^{(t)})\label{eq:connect_2}\\
&=\min\{b_1(\pi^{(t)})+\dots+b_{l'}(\pi^{(t)}),\mathcal{M}\}\label{eq:connect_3}\\
&=\min\{b_1(\pi^{(0)})+\dots+b_{l'}(\pi^{(0)}),\mathcal{M}\}\label{eq:connect_4}\\
&=c_1(\pi^{(0)})+\dots+c_{l'}(\pi^{(0)})\label{eq:connect_5},
\end{align}
where \eqref{eq:connect_1} is by the fact that {\sc CONNECT} may have added 1 to the $h$ value of a node not in $\{\pi^{(t)}(1),\dots,\pi^{(t)}(l)\}$, \eqref{eq:connect_2} is by Claim~\ref{clm:l_prime}, \eqref{eq:connect_3} is by the definition of vector $\mathbf{c}$, \eqref{eq:connect_4} is by the fact that the order of the nodes in the MBR formula does not change the value when $d+1\geq k$ (note that $l'\leq k$) (see the proof of the MBR point of the MFHS scheme).

\par By \eqref{eq:connect_5}, we get a contradiction to the fact that $\mathbf{c}(\pi^{(0)})\succeq \mathbf{h}^{(0)}$ we have initially.

\par \emph{Case~2}: if $1\notin \{\pi^{(t)}(1),\dots,\pi^{(t)}(l')\}$. Let $v=|\{\pi^{(t)}(1),\dots,\pi^{(t)}(l')\}\cap \overline{D}(1)|$. We have that
\begin{align}
h_{\pi^{(0)}(1)}^{(0)}+\dots+&h_{\pi^{(0)}(l')}^{(0)}+h_1^{(0)}\nonumber\\
&\geq h_{\pi^{(t)}(1)}^{(0)}+\dots+h_{\pi^{(t)}(l')}^{(0)}+h_1^{(0)}\label{eq:connect_2_1}\\
&\geq h_{\pi^{(t)}(1)}^{(t)}+\dots+h_{\pi^{(t)}(l')}^{(t)}+(d-v) \label{eq:connect_2_2}\\
&>c_1(\pi^{(t)})+\dots+c_{l'}(\pi^{(t)})+(d-v)\label{eq:connect_2_3},
\end{align}
where \eqref{eq:connect_2_1} is by the fact that $h_{\pi^{(0)}(i)}^{(0)}\geq h_{\pi^{(0)}(i+1)}^{(0)}$, \eqref{eq:connect_2_2} follows by the fact that all the helper nodes in $\{\pi^{(t)}(l'+1),\dots,\pi^{(t)}(n)\}\cap D^{(0)}$ have been incremented by 1 so far or otherwise helper node $l$ would not have been picked by {\sc CONNECT} to be incremented. This means that at least $(d-v)$ nodes not in $\{\pi^{(t)}(1),\dots,\pi^{(t)}(l')\}$ have been incremented by 1. Now, we have two sub-cases.

\par \emph{Sub-case 2.1}: if $1\in \{\pi^{(0)}(1),\dots,\pi^{(0)}(l'+1)\}$. Then, we have that 
\begin{align}\label{eq:cs_equal}
c_1(\pi^{(t)})+\dots+c_{l'}(\pi^{(t)})+&(d-v)=\nonumber\\
&c_1(\pi^{(0)})+\dots+c_{l'+1}(\pi^{(0)}).
\end{align}
The reason is the following. By the construction of $\pi^{(0)}$ (made possible by Claim~\ref{clm:swapping}) in Step~1 of {\sc CONNECT}, \emph{we can see that the order of the nodes in the node permutation excluding node 1 does not change}. Only node 1 may move to a later coordinate. This gives us that $\{\pi^{(0)}(1),\dots,\pi^{(0)}(l'+1)\}=\{\pi^{(t)}(1),\dots,\pi^{(t)}(l'),1\}$. Using the fact that the order of the nodes in the permutation does not change the MBR point, in a similar fashion to Case~1, we got \eqref{eq:cs_equal}. Therefore, we get by \eqref{eq:connect_2_3} that $\mathbf{c}(\pi^{(0)})\nsucceq \mathbf{h}^{(0)}$, i.e., a contradiction with the fact that $\mathbf{h}\in \mathcal{H}$.

\par \emph{Sub-case 2.2}: if $1\notin \{\pi^{(0)}(1),\dots,\pi^{(0)}(l'+1)\}$. We have that
\begin{align}\label{eq:cs_equal_2}
c_1(\pi^{(t)})+\dots+c_{l'}(\pi^{(t)})=c_1(\pi^{(0)})+\dots+c_{l'}(\pi^{(0)}),
\end{align}
by the fact that only node 1 moves in permutation $\pi^{(0)}$ as discussed before. Thus, by \eqref{eq:connect_2_3}, we get
\begin{align}
h_{\pi^{(0)}(1)}^{(0)}+\dots+&h_{\pi^{(0)}(l')}^{(0)}+h_1^{(0)}\nonumber\\
&>c_1(\pi^{(0)})+\dots+c_{l'}(\pi^{(0)})+(d-v)\label{eq:connect_sub_2}.
\end{align}
If $h_{\pi^{(0)}(l'+1)}^{(0)}=h_1^{(0)}$, swap nodes 1 and $\pi^{(0)}(l+1)$ in $\pi^{(0)}$ to get a new permutaiton $\pi'$, and then using \eqref{eq:connect_sub_2} we get
\begin{align}
h_{\pi^{(0)}(1)}^{(0)}+\dots+&h_{\pi^{(0)}(l')}^{(0)}+h_1^{(0)}\nonumber\\
&>c_1(\pi^{(0)})+\dots+c_{l'}(\pi^{(0)})+(d-v)\nonumber\\
&=c_1(\pi')+\dots+c_{l'+1}(\pi')\label{eq:connect_sub_6},
\end{align}
where we get \eqref{eq:connect_sub_6} by the fact that $\{\pi^{(0)}(1),\dots,\pi^{(0)}(l')\}=\{\pi'(1),\dots,\pi'(l')\}$ and the other fact that $c_{l+1}(\pi')=(d-v)$ since $\{\pi^{(t)}(1),\dots,\pi^{(t)}(l')\}=\{\pi'(1),\dots,\pi'(l')\}$. By Claim~\ref{clm:swapping}, equation \eqref{eq:connect_sub_6} yields a contradiction. Now, $h_{\pi^{(0)}(l'+1)}^{(0)}<h_1^{(0)}$ is not possible since $1\notin \{\pi^{(0)}(1),\dots,\pi^{(0)}(l'+1)\}$. Therefore we are left with the case that $h_{\pi^{(0)}(l'+1)}^{(0)}>h_1^{(0)}$. In this case, we can write
\begin{align}
h_{\pi^{(0)}(1)}^{(0)}+&\dots+h_{\pi^{(0)}(l'+1)}^{(0)}\nonumber\\
&\geq h_{\pi^{(0)}(1)}^{(0)}+\dots+h_{\pi^{(0)}(l')}^{(0)}+h_1^{(0)}+1\nonumber\\
&\geq h_{\pi^{(t)}(1)}^{(0)}+\dots+h_{\pi^{(t)}(l)'}^{(0)}+h_1^{(0)}+1\nonumber\\
&\geq h_{\pi^{(t)}(1)}^{(t)}+\dots+h_{\pi^{(t)}(l')}^{(t)}+(d-v) +1\nonumber\\
&>c_1(\pi^{(t)})+\dots+c_{l'}(\pi^{(t)})+(d-v)+1\nonumber\\
&=c_1(\pi^{(0)})+\dots+c_{l'}(\pi^{(0)})+(d-v)+1\label{eq:connect_sub_3}\\
&\geq c_1(\pi^{(0)})+\dots+c_{l'}(\pi^{(0)})+c_{l'+1}(\pi^{(0)})-1+1\label{eq:connect_sub_4}\\
&=c_1(\pi^{(0)})+\dots+c_{l'+1}(\pi^{(0)})\label{eq:connect_sub_5},
\end{align}
where \eqref{eq:connect_sub_3} is by \eqref{eq:cs_equal_2}, \eqref{eq:connect_sub_4} is by the fact that $(d-v) \geq c_{l'+1}(\pi^{(0)})-1$ since the family size is 2. By \eqref{eq:connect_sub_5}, we get a contradiction again. Hence, the proof of the correctness of {\sc CONNECT} is complete.
\end{IEEEproof}

\section{Exact Repair Linear Code Construction for $(n,k,d,r)=(6,3,2,1)$}
In the following, we present a linear \emph{exact repair} LRRC construction that can achieve the MBR point of the MFHS scheme for $(n,k,d,r)=(6,3,2,1)$. By \cite[Proposition~12]{arxiv_multiple}, we have that $\mathcal{M}=4$ packets and $\alpha=2$ packets for $\beta=1$ packet. Recall that according to the MFHS scheme, we can divide the storage nodes into two complete families, family 1 consisting of $\{1,2,3\}$ and family 2 consisting of $\{4,5,6\}$.

\par \emph{Step~1:} Generate a $(6,4)$ systematic MDS code over a finite field $\mathbb{F}$. Denote the generating matrix of this MDS code by
\begin{align}
\mathbf{G}=\left(\begin{array}{cccccc}
1 & 0 & 0 & 0 & a_1 &\underline{a_1}\\
0 & 1 & 0 & 0 & a_2 &\underline{a_2}\\
0 & 0 & 1 & 0 & b_1 &\underline{b_1}\\
0 & 0 & 0 & 1 & b_2 &\underline{b_2}
\end{array}
\right).
\end{align}

\par \emph{Step~2:} Using the same coding vector notation as in Section~\ref{sec:rand_ex}, we store in nodes $\{1,2,3\}$ the packets corresponding to the coding vectors
\begin{align}
\mathbf{Q}_1=\left(\begin{array}{cc}
1 & 0\\
0 & 1\\
0 & 0\\
0 & 0
\end{array}
\right),
\mathbf{Q}_2=\left(\begin{array}{cc}
0 & 0\\
0 & 0\\
1 & 0\\
0 & 1
\end{array}
\right)
\end{align}
\begin{align}
\mathbf{Q}_3=\left(\begin{array}{cc}
a_1 & \underline{a}_1\\
a_2 & \underline{a}_2\\
b_1 & \underline{b}_1\\
b_2 & \underline{b}_2
\end{array}
\right),
\mathbf{Q}_4=\left(\begin{array}{cc}
a_1 & 0\\
a_2 & 0\\
0 & b_1\\
0 & b_2
\end{array}
\right)
\end{align}
\begin{align}\mathbf{Q}_5=\left(\begin{array}{cc}
\underline{a}_1 & 0\\
\underline{a}_2 & 0\\
0 & \underline{b}_1\\
0 & \underline{b}_2
\end{array}
\right),
\mathbf{Q}_6=\left(\begin{array}{cc}
a_1 +\underline{a}_1& 0\\
a_2 + \underline{a}_2&0\\
0 & b_1+\underline{b}_1\\
0 & b_2+\underline{b}_2
\end{array}
\right).
\end{align}

\par By the above, we have completely described the code construction. We are left with verifying that the above code can protect against any $(n-k)=3$ simultaneous node failures, and that the code is exactly repairable using the MFHS scheme. To that end, observe that any two nodes of family 1, i.e., in $\{1,2,3\}$, have together 4 coding vectors that are full rank. It is not hard to see that we also have that any two nodes of family 2 have 4 coding vectors that are full rank. Since any set of 3 nodes must have two nodes from the same family, any $k=3$ nodes can reconstruct the file by the above arguments. 

\par Now, let us look at the repairability of the code. Suppose we wish to repair node 4 and node 3 is unavialable. We access node 1 and get the packet $(a_1 P_1^{(1)}+a_2P_2^{(1)})$, where $P_j^{(i)}$ is the $j$-th packet in node $i$ (the order of the packets in a node is the same as the order of the coding vectors in the corresponding $Q$ matrix). The packet we got so far is the first packet of node 4. We then access node 2 and get the packet $(b_1 P_1^{(2)}+b_2P_2^{(2)})$, which is the second packet of node 4. We have thus exactly repaired node 4 by accessing nodes 1 and 2 of family 1 and communicating no more than one packet from each as we need to have $\beta=1$.

\par Suppose we now wish to repair node 4 and node 1 is unavailable. We get from node 2 the packet $(b_1 P_1^{(2)}+b_2P_2^{(2)})$ and from node 3 the packet $P_1^{(3)}$. To get the first packet of node 4, node 4 has to do the following calculation: $P_1^{(4)}=P_1^{(3)}-(b_1 P_1^{(2)}+b_2P_2^{(2)})$. The second packet of node 4, packet $P_2^{(4)}$, is exactly the packet $(b_1 P_1^{(2)}+b_2P_2^{(2)})$. Therefore, we have repaired node 4 exactly. In a similar manner as above, we can repair node 4 when node 2 is unavailable. The repair of nodes 5 and 6 given any node unavailability is also similar to the repair procedure described above.

\par Let us suppose now that we wish to repair node 1 while node 5 is unavailable. We access node 4 and get $P_1^{(4)}$ and we access node 6 and get $P_1^{(6)}$. Since these communicated packets are linearly independent, we can reconstruct the two packets of node 1. In a similar fashion we repair node 1 for any other given node unavailability. Also, node 2 can be repaired similarly. The repair of node 3 is not hard, so we leave it for the reader. 

\par We have therefore proved that the above code construction can reconstruct the file from any set of $k=3$ nodes and is exactly repairable. 

\appendices
\section{Proof of Claim~\ref{clm:swapping}}\label{app:swapping_claim}
\par In this appendix we will prove Claim~\ref{clm:swapping}. We begin the proof by showing that for $i>k-1$ Claim~\ref{clm:swapping} is trivially true. Consider $i=k$ first. We have that
\begin{align}
c_1(\pi)+\dots+c_{i-1}(\pi)=c_1(\pi')+\dots+c_{i-1}(\pi')
\end{align}
and since $i=k$, we have that
\begin{align}
c_1(\pi)+\dots+c_i(\pi)=c_1(\pi')+\dots+c_i(\pi')=\mathcal{M}
\end{align}
by the MBR point of the MFHS scheme. By the above facts, we get that $c_i(\pi)=c_i(\pi')$ which implies that $\mathbf{c}(\pi)=\mathbf{c}(\pi')$. Then, $\mathbf{c}(\pi)\succeq \mathbf{h}$ implies $\mathbf{c}(\pi')\succeq \mathbf{h}$ directly.

\par Now, since $c_i(\pi)$ is 0 for $i>k$ for any permutation $\pi$ due to truncation, then for $i>k$, we have that $\mathbf{c}(\pi)=\mathbf{c}(\pi')$ and again $\mathbf{c}(\pi)\succeq \mathbf{h}$ directly implies $\mathbf{c}(\pi')\succeq \mathbf{h}$.

\par Therefore, we only need to consider the case when $i\leq k-1$. We divide the proof into two cases:
\par \emph{Case~1}: $h_{\pi(1)}+\dots+h_{\pi(i)}=c_1(\pi)+\dots+c_i(\pi)$. Since $\mathbf{c}(\pi)\succeq \mathbf{h}$, we have that $h_{\pi(i+1)}\leq c_{i+1}(\pi)$. Recall that $\mathbf{c}(\pi)$ is the truncated version of vector $\mathbf{b}(\pi)$. If $FI(\pi(i))=FI(\pi(i+1))$, we have that $b_i(\pi)=b_{i+1}(\pi)=b_i(\pi')=b_{i+1}(\pi')$. Thus, this case is trivial.

\par If $FI(\pi(i))\neq FI(\pi(i+1))$, we have that $b_i(\pi')\geq b_{i+1}(\pi)$ (by the definition of function $z_i$). Now, if $b_i(\pi')=c_i(\pi')$, then we have that $b_i(\pi')=c_i(\pi')\geq c_{i+1}(\pi)$. If on the other hand, $b_i(\pi')>c_i(\pi')$, since $\sum_{m=1}^{i-1}b_m(\pi)=\sum_{m=1}^{i-1}b_m(\pi')$, then $c_i(\pi')\geq c_i(\pi)\geq c_{i+1}(\pi)$ where the first inequality follows by the fact that $c_i(\pi)>c_i(\pi')$ is not possible since that gives $\sum_{m=1}^n c_m >\mathcal{M}$ yielding a contradiction, and the second inequality follows by Lemma~\ref{lem:b_c}. By the above arguments, we get that $c_i(\pi')\geq c_{i+1}(\pi)$. Thus, we get that $h_{\pi(i+1)}\leq c_{i+1}(\pi)\leq c_i(\pi')$. Therefore,
\begin{align*}
h_{\pi(1)}+\dots+h_{\pi(i)}&=h_{\pi'(1)}+\dots+h_{\pi'(i)}\\
&\leq c_1(\pi')+\dots+c_i(\pi').
\end{align*}

\par Now, we have that $\pi'(i+1)=\pi(i)$. Originally, we had 
\begin{align*}
h_{\pi(1)}+\dots+h_{\pi(i+1)}\leq c_1(\pi)+\dots+c_{i+1}(\pi).
\end{align*}
Since 
\begin{align*}
b_1(\pi)+\dots+b_{i+1}(\pi)=b_1(\pi')+\dots+b_{i+1}(\pi'),
\end{align*}
as we saw in the proof of the MBR point in \cite[Proposition~7]{arxiv1} when $d+1\geq k$, we get
\begin{align*}
\min\{b_1(\pi)+\dots+&b_{i+1}(\pi),\mathcal{M}\}=\\
&\min\{b_1(\pi')+\dots+b_{i+1}(\pi'),\mathcal{M}\}.
\end{align*}
This implies that 
\begin{align*}
c_1(\pi)+\dots+c_{i+1}(\pi)=c_1(\pi')+\dots+c_{i+1}(\pi').
\end{align*}
Therefore, 
\begin{align*}
h_{\pi(1)}+\dots+h_{\pi(i+1)}\leq c_1(\pi')+\dots+c_{i+1}(\pi').
\end{align*}
Hence, $\mathbf{c}(\pi')\succeq \mathbf{h}$ for Case~1.

\par \emph{Case~2}: $h_{\pi(1)}+\dots+h_{\pi(i)}<c_1(\pi)+\dots+c_i(\pi)$. We have $b_i(\pi')\geq b_{i+1}(\pi')\geq b_i(\pi)-1$, where the first inequality follows by Lemma~\ref{lem:b_c} and the second inequality by the fact that $FI(\pi(i))\neq FI(\pi(i+1))$. We now want to show that this implies that $c_i(\pi')\geq c_i(\pi)-1$.
\begin{itemize}
\item If $b_{i-1}(\pi)>c_{i-1}(\pi)$, then $c_i(\pi)=0$ and we thus trivially have $c_i(\pi')\geq c_i(\pi)-1$.
\item If $b_i(\pi)>c_i(\pi)$, then $b_i(\pi')\geq b_i(\pi)-1\geq c_i(\pi)$. Since $\sum_{m=1}^{i-1} b_m(\pi)=\sum_{m=1}^{i-1} b_m(\pi')$, $c_i(\pi')=c_i(\pi)\geq c_i(\pi)-1$.
\item If $b_{i+1}(\pi)>c_{i+1}(\pi)$, then we have $b_i(\pi)=c_i(\pi)$.
\begin{itemize}
\item If $b_i(\pi')=b_i(\pi)+1$, then $c_i(\pi)\leq c_i(\pi')$.
\item If $b_i(\pi')=b_i(\pi)$, then $c_i(\pi')=c_i(\pi)$.
\item If $b_i(\pi')=b_i(\pi)-1$, then $c_i(\pi')=c_i(\pi)-1$.
\end{itemize} 
\end{itemize}
Therefore, we indeed have that $c_i(\pi')\geq c_i(\pi)-1$.

\par By this fact, we have that
\begin{align}
h_{\pi(1)}+\dots+h_{\pi(i)}&=h_{\pi'(1)}+\dots+h_{\pi'(i)}\\
&\leq c_1(\pi')+\dots+c_i(\pi').
\end{align}
Now, by the same argument as in Case~1, we get that
\begin{align}
h_{\pi(1)}+\dots+h_{\pi(i+1)}\leq c_1(\pi)+\dots+c_{i+1}(\pi).
\end{align}
Therefore, we get that $\mathbf{c}(\pi')\succeq \mathbf{h}$ for Case~2 too. Hence, the proof of this claim is complete.

\bibliography{paper}
\bibliographystyle{IEEEtranS}

\end{document}